\documentclass[journal]{IEEEtran}

\usepackage{epsfig,amsmath,amssymb,epsf,amsthm,scalefnt,multirow,subfig}
\usepackage{xcolor}
\usepackage{float}
\usepackage{cite}
\usepackage{psfrag}
\usepackage{graphics,amsmath,amssymb,graphicx,amsthm,scalefnt,multirow,dsfont,subfig}

\usepackage{epsfig,amsmath,amssymb,epsf,amsthm,scalefnt,multirow}
\usepackage{xcolor}
\usepackage{float}
\usepackage{cite}
\usepackage{psfrag}
\usepackage{cleveref}
\usepackage{mathrsfs}
\usepackage{mathtools}
\usepackage{algpseudocode}
\usepackage{algorithm}
\usepackage{enumerate}
\usepackage{stfloats}
\usepackage{amsmath}
\usepackage{tabularx}

\newtheorem*{lemma*}{Lemma}

  \def\cC{{\mathcal{C}}}

 \def\cN{{\mathcal{N}}} \def\cO{{\mathcal{O}}}

\def\diag{\mathop{\mathrm{diag}}}

\def\trace{\mathop{\mathrm{tr}}}

 \def\bgamma{{\pmb{\gamma}}}

\def\b0{{\pmb{0}}}\def\bLambda{{\pmb{\Lambda}}} 

\def\ba{{\mathbf{a}}}  \def\bc{{\mathbf{c}}}

\def\bm{{\mathbf{m}}} \def\bn{{\mathbf{n}}}  
  \def\bs{{\mathbf{s}}} 
   \def\bx{{\mathbf{x}}}
\def\by{{\mathbf{y}}}   

\def\bA{{\mathbf{A}}}  \def\bC{{\mathbf{C}}} 
  \def\bG{{\mathbf{G}}} 
\def\bI{{\mathbf{I}}}   
 \def\bN{{\mathbf{N}}}  
  \def\bS{{\mathbf{S}}} 
  \def\bW{{\mathbf{W}}} 
\def\bY{{\mathbf{Y}}}

\begin{document}	
\bstctlcite{IEEEexample:BSTcontrol}
\title{A Bayesian Framework For Cascaded Channel Estimation in RIS-Aided mmWave Systems}

\author{\IEEEauthorblockN{Gyoseung Lee and Junil Choi}
	\thanks{This work was partly supported by the Institute of Information \& Communications Technology Planning \& Evaluation (IITP)-ITRC (Information Technology Research Center) grant funded by the Korea government (MSIT) (IITP-2025-RS-2020-II201787, contribution rate 50\%) and by IITP grant funded by the Korea government (MSIT) (No. RS-2024-00395824, Development of Cloud virtualized RAN (vRAN) system supporting upper-midband).}
	\thanks{Gyoseung Lee and Junil Choi are with the School of Electrical Engineering, Korea Advanced Institute of Science and Technology, Daejeon 34141, South Korea (e-mail: \{iee4432; junil\}@kaist.ac.kr).}
	}
\maketitle

\begin{abstract}
In this paper, we investigate cascaded channel estimation for reconfigurable intelligent surface (RIS)-aided millimeter-wave multi-user communication systems.
Since the complex channel gains of the cascaded RIS channel are generally non-Gaussian,
the use of the linear minimum mean squared error (LMMSE) estimator leads to inevitable performance degradation.
To tackle this issue, we propose a variational inference-based framework that approximates the complex channel gains using a complex adaptive Laplace prior, which effectively captures their probability distributions in a tractable way.
Numerical results demonstrate that the proposed estimator outperforms conventional estimators including least squares and LMMSE in terms of cascaded channel estimation error.
\end{abstract}


\section{Introduction}\label{sec1}
To achieve high data rates required to meet diverse demands of beyond 5G communications, millimeter-wave (mmWave) systems are considered as a promising solution thanks to the availability of large bandwidths in the previously unused spectrum between 30 and 300 GHz \cite{Wang:2018}.
However, mmWave systems are highly sensitive to channel variations and blockages due to the high frequency propagation characteristics, which result in severe path-loss and high penetration loss, thus making it challenging to effectively utilize mmWave spectra in practical wireless communication applications.
To address this challenge, various metasurface-based technologies have attracted significant attention in both academia and industry, including reconfigurable intelligent surfaces (RISs), electromagnetic surfaces composed of passive, reconfigurable units capable of adjusting electromagnetic properties of incoming signals \cite{Wu:2020, Basar:2019}, and holographic multiple-input multiple-output (MIMO), which aims to overcome hardware limitations in conventional MIMO systems \cite{Li:2024, Li:2025}.
Focusing on the RIS, in cases where potential blockages between a base station (BS) and user equipment (UE) are present, an RIS can provide a virtual line-of-sight link, thereby enhancing coverage and preserving reliability in mmWave~systems.

To fully achieve potential advantages of the RIS, the BS or UE must acquire channel state information (CSI) to align RIS phase shifts with wireless channels. Generally, an RIS consists entirely of passive elements without radio frequency (RF) chains, which results in the BS or UE observing a cascaded UE-RIS-BS channel and makes it challenging to apply classical channel estimation techniques developed without RISs.

In mmWave systems, due to the limited scattering environments, wireless channels are typically modeled using a double-directional channel model that describes the channels based on individual propagation paths \cite{Li:2014}. Under this type of model, several papers have focused on estimating parameters of RIS-related channels associated with dominant paths such as angles of arrival (AoAs) or angles of departure (AoDs) \cite{Wang:2023, Zhang:2021, Zhou:2024}. 
In \cite{Wang:2023}, a hierarchical beam training technique was developed to estimate angles between transceivers.
The work in \cite{Zhang:2021} developed an angle estimation algorithm for the RIS-related channels based on the estimation of signal parameter via rotational invariance technique (ESPRIT) and multiple signal classification (MUSIC).
In \cite{Zhou:2024}, angle estimation methods based on orthogonal matching pursuit (OMP) and atomic norm minimization (ANM) were developed for the cases where monostatic and bistatic full-duplex BSs are exploited.
However, in such works, the search for techniques related to complex path gain estimation of the cascaded channel remains rather elusive.

Typically, it is reasonable to model the complex path gains in wireless channels using a Gaussian distribution, and well-known approaches for estimating these gains include lease squares (LS) and linear minimum mean squared error (LMMSE) estimators. 
However, in the cascaded channel, the complex path gains are products of the separate RIS-related channels, which leads to inevitable performance degradation for the LMMSE estimator since the product of two independent Gaussian random variables is no longer Gaussian.

In this paper, to tackle the aforementioned issue, we propose a Bayesian framework to estimate the complex path gains of the cascaded channel in RIS-aided mmWave communication systems. For tractable posterior inference, we approximate the distribution of the path gains using a complex adaptive Laplace distribution, which effectively approximates the shape of the true distribution that is the product of two independent zero-mean Gaussian random variables. Based on this model, we employ variational inference (VI) to derive approximate posterior distributions. Our numerical results verify that in terms of the cascaded channel estimation accuracy, the proposed estimator outperforms conventional estimators including the LS and LMMSE estimators.

The rest of the paper is organized as follows.
Section \ref{sec2} presents the system model for the assumed RIS-aided multi-UE system.
In Section \ref{sec3}, the problem formulation for estimating the complex path gains of the cascaded channel is investigated. The proposed estimator is developed in Section \ref{sec4}, and simulations are carried out in Section \ref{sec_numerical}.
Finally, we conclude the paper in Section \ref{conclusion}.

\section{System Model}\label{sec2}
\subsection{Signal model}

We investigate an uplink multi-UE system, which consists of a BS with $N$ antennas in a uniform linear array (ULA) and $K$ UEs equipped with a single transmit antenna. In this system, a single RIS composed of $L$ purely passive elements in a uniform planar array (UPA) is present, whose configurations are controlled by the BS via a low-rate network connection.

As in \cite{Zhang:2021}, we assume that the direct link channels between the BS and UEs are completely blocked.
Assuming all UEs transmit signals with power $P$, the uplink signal received by the BS at time instant $t$ is given~by
\begin{align} \label{y_t}
	\by[t] &= \sqrt{P} \sum_{k=1}^K \bG\diag(\bs[t])\mathbf{f}_k x_k[t] + \bn_{\mathrm{B}}[t],
\end{align}
where $x_k[t] \in \mathbb{C}$ is the signal sent from the $k$-th UE satisfying $\mathbb{E}[\vert x_k[t] \vert^2] =1$, and $\bn_{\mathrm{B}}[t] \sim \cC\cN(\bn_{\mathrm{B}}[t] \vert \b0_N, \sigma_{\mathrm{B}}^2 \bI_N)$ is additive white Gaussian noise (AWGN) at the BS with variance $\sigma_{\mathrm{B}}^2$. The uplink channel from the RIS to the BS is denoted by  $\bG \in \mathbb{C}^{N \times L}$, and the uplink channel from the $k$-th UE to the RIS is represented by $\mathbf{f}_k \in \mathbb{C}^{L \times 1}$. The vector of RIS passive reflection coefficients is $\bs[t] = [s_1[t],\cdots,s_L[t]]^{\mathrm{T}} \in \mathbb{C}^{L \times 1}$, with reflection amplitudes $\vert s_{\ell}[t] \vert = 1$ and phase shifts $\angle s_{\ell}[t] \in [0, 2\pi)$.

For the proposed channel estimation framework, the following transmission protocol is considered: a coherence block of length $T_{\mathrm{p}}=T\tau$ is divided into $T$ subblocks, each consisting of $\tau$ time slots. Within each subblock, the RIS configuration remains fixed, and all UEs transmit the identical pilot sequences across all $T$ subblocks. Then, the signal received by the BS in the $u$-th time slot of the $t$-th subblock $\by[t,u]=\by[(t-1)\tau+u]$ is expressed as
\begin{align} \label{y_t_u}
	\by[t,u] &= \sqrt{P} \sum_{k=1}^K \bG\diag(\bs[t])\mathbf{f}_k x_k[u] + \bn_{\mathrm{B}}[t,u].
\end{align}
The measurement matrix at the BS obtained by stacking (\ref{y_t_u}) over $\tau$ time slots for the $t$-th subblock is given by
\begin{align} \label{Y_t}
	\bY[t] &= [\by[t,1],\cdots,\by[t,\tau]] \nonumber
	\\ &= \sqrt{P}\sum_{k=1}^K \bG \diag(\bs[t]) \mathbf{f}_k \bx_k^{\mathrm{T}} + \bN_{\mathrm{B}}[t],
\end{align}
where $\bx_k=[x_k[1],\cdots,$ $x_k[\tau]]^{\mathrm{T}} \in \mathbb{C}^{\tau \times 1}$, and $\bN_{\mathrm{B}}[t] = [\bn_{\mathrm{B}}[t,1],\cdots,\bn_{\mathrm{B}}[t,\tau]] \in \mathbb{C}^{N \times \tau}$. Note that, we assume that all UEs transmit orthogonal pilot sequences such that $\bx_k^{\mathrm{T}}\bx_g^* = \tau$ for $k=g$ and $\bx_k^{\mathrm{T}}\bx_g^* = 0$ for $k \neq g $, which enables to the separation of the measurement signal for the $k$-th UE from (\ref{Y_t}) given by
\begin{align} \label{Y_t_separation}
	\frac{1}{\tau} \bY[t] \bx_k^* &= \sqrt{P}\bG\diag(\bs[t])\mathbf{f}_k + \frac{1}{\tau}\bN_{\mathrm{B}}[t] \bx_k^*
	\nonumber \\ &= \sqrt{P}\bG\diag(\mathbf{f}_k)\bs[t] + \frac{1}{\tau}\bN_{\mathrm{B}}[t] \bx_k^*.
\end{align}
Finally, the overall measurement matrix $\bY_k \in \mathbb{C}^{N \times T}$ obtained by collecting all $T$ subblocks in (\ref{Y_t_separation}) is given by
\begin{align} \label{Y_k}
	\bY_k &= \frac{1}{\tau}[\bY[1]\bx_k^* , \cdots , \bY[T]\bx_k^*] \nonumber \\& = \sqrt{P} \bG \diag({\mathbf{f}_k}) \bS + \bN_k,
\end{align}
where $\bS=[\bs[1],\cdots,\bs[T]] \in \mathbb{C}^{L \times T}$, and $\bN_k = \frac{1}{\tau}[\bN_{\mathrm{B}}[1]\bx_k^*, \cdots, \bN_{\mathrm{B}}[T]\bx_k^*] \in \mathbb{C}^{N \times T}$.

\subsection{Channel model} \label{sec2_2}
In mmWave systems, wireless channels are typically modeled as a geometric channel model \cite{Li:2014, Geometric_1, Zhang:2021}, due to the limited scattering environment in mmWave spectra. Under this model, the uplink RIS-BS channel $\bG$ is given by
\begin{align} \label{G}
	\bG &= \sqrt{\frac{NL}{M_{\mathrm{RB}}}} \sum_{m=1}^{M_{\mathrm{RB}}} \alpha_{\mathrm{RB},m} \ba_{\mathrm{B}}(\phi_{m}) \ba_{\mathrm{R}}^{\mathrm{H}}(\theta_{\mathrm{RB},m}^{\mathrm{Azi}}, \theta_{\mathrm{RB},m}^{\mathrm{Ele}}),
\end{align}
where $M_{\mathrm{RB}}$ denotes the number of propagation paths for this channel, and $\alpha_{\mathrm{RB},m} \sim \cC\cN(\alpha_{\mathrm{RB},m} \vert 0, \sigma_{\mathrm{RB}}^2)$ is the complex gain of the $m$-th path, which is independent and identically distributed (i.i.d.) with zero mean and variance $\sigma_{\mathrm{RB}}^2$ that depends on the path-loss. The array steering vectors at the BS and RIS, denoted by $\ba_{\mathrm{B}}(\cdot) \in \mathbb{C}^{N\times 1}$ and $\ba_{\mathrm{R}}(\cdot) \in \mathbb{C}^{L \times 1}$, respectively, are defined similarly as in \cite{Zhang:2021}, and $\phi_{m}$ is the AoA of the $m$-th path, and $\theta_{\mathrm{RB},m}^{\mathrm{Azi}}$ and $\theta_{\mathrm{RB},m}^{\mathrm{Ele}}$ are the azimuth and elevation AoDs of the $m$-th path, respectively.
For simplicity, we reformulate $\bG$ in (\ref{G}) as
\begin{align} \label{G_simple}
	\bG = \bA_{\mathrm{B,RB}} \diag(\boldsymbol{\alpha}_{\mathrm{RB}}) \bA_{\mathrm{R,RB}}^{\mathrm{H}},
\end{align}
where $\bA_{\mathrm{B,RB}}=[\ba_{\mathrm{B}}(\phi_{1}),\cdots,\ba_{\mathrm{B}}(\phi_{M_{\mathrm{RB}}})] \in \mathbb{C}^{N \times M_{\mathrm{RB}}}$, $\bA_{\mathrm{R,RB}}=[\ba_{\mathrm{R}}(\theta_{\mathrm{RB},1}^{\mathrm{Azi}}, \theta_{\mathrm{RB},1}^{\mathrm{Ele}}),\cdots,\ba_{\mathrm{R}}(\theta_{\mathrm{RB},M_{\mathrm{RB}}}^{\mathrm{Azi}}, \theta_{\mathrm{RB},M_{\mathrm{RB}}}^{\mathrm{Ele}})] \in \mathbb{C}^{L \times M_{\mathrm{RB}}}$, and $\boldsymbol{\alpha}_{\mathrm{RB}}=\sqrt{\frac{NL}{M_{\mathrm{RB}}}}[\alpha_{\mathrm{RB},1},\cdots,\alpha_{\mathrm{RB},M_{\mathrm{RB}}}]^{\mathrm{T}} \in \mathbb{C}^{M_{\mathrm{RB}}\times 1}$.

Similarly, the channel from the $k$-th UE to the RIS $\mathbf{f}_k$ is expressed~as
\begin{align} \label{f_k}
	\mathbf{f}_k &= \sqrt{\frac{L}{M_{\mathrm{UR},k}}} \sum_{m=1}^{M_{\mathrm{UR},k}} \alpha_{\mathrm{UR},k,m} \ba_{\mathrm{R}}(\theta_{\mathrm{UR},k,m}^{\mathrm{Azi}}, \theta_{\mathrm{UR},k,m}^{\mathrm{Ele}}),
\end{align}
where $M_{\mathrm{UR},k}$ is the number of propagation paths for this channel, $\alpha_{\mathrm{UR},k,m} \sim \cC\cN(\alpha_{\mathrm{UR},k,m} \vert 0, \sigma_{\mathrm{UR},k}^2)$ is the i.i.d. complex gain of the $m$-th path with zero mean and variance $\sigma_{\mathrm{UR},k}^2$, and $\theta_{\mathrm{UR},k,m}^{\mathrm{Azi}}$ and $\theta_{\mathrm{UR},k,m}^{\mathrm{Ele}}$ respectively denote the azimuth and elevation AoAs of the $m$-th path.
Based on (\ref{f_k}), $\mathbf{f}_k$ can be reformulated~as 
\begin{align} \label{f_k_simple}
	\mathbf{f}_k = \bA_{\mathrm{R,UR},k} \boldsymbol{\alpha}_{\mathrm{UR},k},
\end{align}
where $\bA_{\mathrm{R,UR},k}=[\ba_{\mathrm{R}}(\theta_{\mathrm{UR},k,1}^{\mathrm{Azi}}, \theta_{\mathrm{UR},k,1}^{\mathrm{Ele}}),\cdots,\ba_{\mathrm{R}}(\theta_{\mathrm{UR},k,M_{\mathrm{UR},k}}^{\mathrm{Azi}},$ $\theta_{\mathrm{UR},k,M_{\mathrm{UR},k}}^{\mathrm{Ele}})]$ $\in \mathbb{C}^{L \times M_{\mathrm{UR},k}}$, and $\boldsymbol{\alpha}_{\mathrm{UR},k}=\sqrt{\frac{L}{M_{\mathrm{UR},k}}}[\alpha_{\mathrm{UR},k,1},$ $\cdots,\alpha_{\mathrm{UR},k,M_{\mathrm{UR},k}}]^{\mathrm{T}} \in \mathbb{C}^{M_{\mathrm{UR},k}\times 1}$.

\section{Problem Formulation} \label{sec3}
Let $\bC_k =\bG \diag({\mathbf{f}_k})=\bG \diamond \mathbf{f}_k^{\mathrm{T}}$ be the cascaded channel between the BS and the $k$-th UE, where $\diamond$ denotes the Khatri-Rao product. Our goal is to develop an estimator for $\bC_k$ from $\bY_k$ in (\ref{Y_k}), based on which $\by_k = \mathrm{vec}(\bY_k)$ is given by
\begin{align} \label{y_k}
	\by_k &= \sqrt{P}(\bS^{\mathrm{T}} \otimes \bI_N)\mathrm{vec}(\bG \diamond \mathbf{f}_k^{\mathrm{T}}) + \mathrm{vec}(\bN_k) \nonumber \\ & = \bar{\bS} \bc_k + \bn_k,
\end{align}
where $\otimes$ denotes the Kronecker product, $\bar{\bS} = \sqrt{P}(\bS^{\mathrm{T}} \otimes \bI_N)$, and $\bc_k = \mathrm{vec}(\bG \diamond \mathbf{f}_k^{\mathrm{T}})$ is the vectorized cascaded channel for the $k$-th UE.
As demonstrated in \cite{Gyoseung:2024}, $\bc_k$ can be simplified as
\begin{equation} \label{c_k_simple}
	\bc_k = \bW_k \boldsymbol{\alpha}_k,
\end{equation}
where $\bW_k = \left(\left(\bA_{\mathrm{R,RB}}^{\mathrm{H}} \diamond \bA_{\mathrm{R,UR},k}^{\mathrm{T}}\right)^{\mathrm{T}} \diamond \tilde{\bA}_{\mathrm{B,RB},k}\right)$ with $\tilde{\bA}_{\mathrm{B,RB},k} = \bA_{\mathrm{B,RB}} \left(\bI_{M_{\mathrm{RB}}} \otimes \boldsymbol{1}_{M_{\mathrm{UR},k}}^{\mathrm{T}} \right)$, and $\boldsymbol{\alpha}_k = \boldsymbol{\alpha}_{\mathrm{RB}} \otimes \boldsymbol{\alpha}_{\mathrm{UR},k}$.

From (\ref{c_k_simple}), we see that $\bc_k$ can be decomposed into the following two parts: a matrix $\bW_k$ that contains the AoAs/AoDs of the RIS-related channels, and a vector $\boldsymbol{\alpha}_k$ whose entries are products of the i.i.d. complex path gains associated with the RIS-related links.
Note that, the AoAs/AoDs vary much slower than the complex path gains and can be assumed to remain fixed across multiple channel coherence blocks \cite{Geometric_1, Kim:2024}, implying that these angles can be accurately estimated over a long period.
In practice, the AoAs/AoDs of the RIS-related channels can be estimated using the methods developed in \cite{Wang:2023, Zhang:2021, Zhou:2024}.


Based on the above discussion, assuming that the angles of the RIS-related channels are known at the BS in advance, our goal boils down to estimate $\boldsymbol{\alpha}_k$ from $\by_k$. From (\ref{c_k_simple}), it is observed that each element in $\boldsymbol{\alpha}_{k}$ is the product of two i.i.d. zero-mean Gaussian random variables, with variances depending on the path-losses of the RIS-related links.
In general, the resulting distribution of the product of independent Gaussian random variables is not Gaussian distributed \cite{Donoughue:2012}, implying inevitable performance degradation when using the LMMSE estimator.
Specifically, according to \cite{Donoughue:2012}, if $x_1 \sim \cC\cN (0,\sigma_1^2)$ and $x_2 \sim \cC\cN(0, \sigma_2^2)$ are independent, the probability density function (PDF) of $z=x_1 x_2$ is derived by
\begin{align} \label{pdf_z} 
	f(z) = \frac{2 \vert z \vert}{\pi \sigma_1^2 \sigma_2^2} K_0 \left(\frac{2\vert z\vert}{\sigma_1 \sigma_2}\right), 
\end{align}
where $K_{\nu}(\cdot)$ denotes the modified Bessel function of the second kind of order $\nu$.
However, directly addressing the form of (\ref{pdf_z}) for our estimation problem is challenging due to the complicated structure of $K_0(\cdot)$.
To address this issue, in the following subsection we will approximate the distribution of $\boldsymbol{\alpha}_k$ using a complex adaptive Laplace distribution, a variant of the Laplace distribution tailored for complex-valued signals that has a similar probability density to (\ref{pdf_z}) due to its exponentially decaying behavior\footnote{From the numerical analysis, we verified that our choice of the complex adaptive Laplace distribution well approximates the true distribution of the complex channel gains of the cascaded channel.}. 
Based on this approximation, we aim to design an estimator that performs approximate posterior inference on $\boldsymbol{\alpha}_k$ from $\by_k$ minimizing the channel estimation error between $\bc_k$ and its estimate.

\section{Proposed Bayesian Estimator} \label{sec4}
In this section, we propose a Bayesian estimator to perform approximate posterior inference on $\boldsymbol{\alpha}_k$ from $\by_k$, where we adopt VI with the mean-field approximation to derive approximate posterior distributions of the variables related to $\boldsymbol{\alpha}_k$.
We first introduce a hierarchical Bayesian model, which enables tractable posterior inference for the considered problem, and then derive the approximate posterior distributions for all random variables using the VI~approach.

\subsection{Hierarchical Bayesian model}
Based on (\ref{y_k}) and (\ref{c_k_simple}), the conditional distribution of $\by_k$~is
\begin{align} \label{prob_y_k}
	p(\by_k \vert \boldsymbol{\alpha}_{k},\beta) &= \cC\cN(\by_k \vert \bS_{\bc,k}\boldsymbol{\alpha}_{k}, \beta^{-1}\bI_{NT} ),
\end{align}
where $\bS_{\bc,k}= \bar{\bS} \bW_k$, and $\beta = 1/\sigma_{\mathrm{B}}^2$ is the inverse of unknown noise variance at the BS $\sigma_{\mathrm{B}}^2$ that is generally unknown in practice, assumed to be gamma distributed as
\begin{align} \label{prob_beta}
	p(\beta) &=\mathrm{Gamma}(\beta \vert a,b) = \frac{b^a}{\Gamma(a)}\beta^{a-1}\exp(-b\beta).
\end{align}
Here, $\Gamma(\cdot)$ is the gamma function, and $a$ and $b$ are respectively the shape and rate parameters to be chosen. Note that it is generally useful to set $a$ and $b$ to sufficiently small values to ensure a broad hyperprior \cite{Tipping:2001}.

To model the complex adaptive Laplace distribution for the prior over $\boldsymbol{\alpha}_k$, we introduce the following random variables: $\boldsymbol{\lambda} =[\lambda_1,\cdots,\lambda_{M_k}]^{\mathrm{T}} \in \mathbb{C}^{M_k \times 1}$ and $\bgamma = [\gamma_1,\cdots,\gamma_{M_k}]^{\mathrm{T}} \in \mathbb{C}^{M_k \times 1}$, where $M_k = M_{\mathrm{UR},k} M_{\mathrm{RB}}$.
Based on this, $\boldsymbol{\alpha}_k$ conditioned on $\boldsymbol{\lambda}$ is assumed to be Gaussian distributed given by
\begin{align} \label{prob_alpha_k}
	p(\boldsymbol{\alpha}_{k} \vert \boldsymbol{\lambda}) &= \cC\cN(\boldsymbol{\alpha}_{k} \vert \b0_{M_k},\bLambda),
\end{align}
where $\bLambda=\diag(\boldsymbol{\lambda})$. The PDF of $\boldsymbol{\lambda}$ conditioned on $\bgamma$ is modeled as the following gamma distribution given by
\begin{align} \label{prob_lambda}
	p(\boldsymbol{\lambda} \vert \bgamma)  &= \prod_{i=1}^{M_k} \mathrm{Gamma}\left( \lambda_i \middle\vert \frac{3}{2}, \frac{\gamma_i}{4} \right).
\end{align}
Note that the shape and rate parameters in (\ref{prob_lambda}) are necessary to construct the prior defined in (\ref{prob_marginal}). Finally, $\bgamma$ is modeled~as
\begin{align} \label{prob_gamma}
	p(\bgamma) &= \prod_{i=1}^{M_k}\mathrm{Gamma}(\gamma_i \vert a, b).
\end{align}
From (\ref{prob_alpha_k}) and (\ref{prob_lambda}), the marginal distribution $p(\boldsymbol{\alpha}_{k} \vert \bgamma) = \int p(\boldsymbol{\alpha}_k \vert \boldsymbol{\lambda}) p(\boldsymbol{\lambda} \vert \bgamma) \mathrm{d} \boldsymbol{\lambda}$ is computed by
\begin{align} \label{prob_marginal}
	p(\boldsymbol{\alpha}_{k} \vert \bgamma) = \frac{\prod_{i=1}^{M_k}\gamma_i}{(2\pi)^{M_k}}\exp\left( - \sum_{i=1}^{M_k} \sqrt{\gamma_i} \vert \alpha_{k,i} \vert \right),
\end{align}
which corresponds to the complex adaptive Laplace prior \cite{Bai:2023}.

\subsection{Proposed VI-based estimator}
Let $\Omega=\{\boldsymbol{\alpha}_k, \boldsymbol{\lambda}, \bgamma, \beta\}$ be the set of hidden variables defined in the previous subsection. Our goal is to infer the posterior distribution $p(\Omega \vert \by_k)$, which, however, is intractable due to the multi-dimensional integration involved in its computation. 
To tackle this issue, we propose a VI-based estimator, which enables approximate posterior computation and facilitates a tractable posterior analysis.

The goal of VI is to derive a variational distribution $q(\Omega)$ that approximates the true posterior by minimizing the Kullback-Leibler (KL) divergence between $q(\Omega)$ and $p(\Omega \vert \by_k)$. Using the mean-field theory \cite{Tipping:2001}, $q(\Omega)$ can be factorized as
\begin{align} \label{mean_field}
	q(\Omega) = q(\boldsymbol{\alpha}_k) q(\boldsymbol{\lambda}) q(\bgamma) q(\beta).
\end{align}
Based on the factorized constraint in (\ref{mean_field}), the optimal approximate posterior minimizing the KL divergence is \cite{Bishop:2006}
\begin{align} \label{VI_solution}
	\log q^{\star}(\Omega_i) \propto  \underbrace{\mathbb{E}_{\prod_{j \neq i}q(\Omega_j)} [\log p(\Omega, \by_k)]}_{\triangleq \langle \log p(\Omega, \by_k) \rangle_{-\Omega_i}}, \enspace \forall i.
\end{align}
Here, $\Omega_i$ denotes the $i$-th element in $\Omega$, and $\langle \cdot \rangle_{-\Omega_i}$ indicates the expectation with respect to $\prod_{j \neq i} q(\Omega_j)$. For convenience, in the following we will use $\langle \cdot \rangle$ to denote the expectation with respect to all variables in $\Omega$.

\textbf{1. Derivation of $q(\boldsymbol{\alpha}_{k})$}: First, plugging in (\ref{prob_y_k}) and (\ref{prob_alpha_k}) to (\ref{VI_solution}) leads to
\begin{align}
	\log q(\boldsymbol{\alpha}_{k}) &\propto \langle \log p(\by_k \vert \boldsymbol{\alpha}_{k},\beta)+\log p(\boldsymbol{\alpha}_{k} \vert \boldsymbol{\lambda}) \rangle_{-\boldsymbol{\alpha}_k} \nonumber
	\\ & \propto -(\boldsymbol{\alpha}_{k}-\bm_{\boldsymbol{\alpha}_{k}})^{\mathrm{H}}\bC_{\boldsymbol{\alpha}_{k}}^{-1}(\boldsymbol{\alpha}_{k}-\bm_{\boldsymbol{\alpha}_{k}}),
\end{align}
which implies that $q(\boldsymbol{\alpha}_k)$ follows a Gaussian distribution with
\begin{align}
	\bm_{\boldsymbol{\alpha}_{k}} &= \langle \beta \rangle \bC_{\boldsymbol{\alpha}_{k}} \bS_{\bc,k}^{\mathrm{H}}\by_k, \label{posterior_mean_alpha_k}
	\\ \bC_{\boldsymbol{\alpha}_{k}} &= \left(\langle \beta \rangle \bS_{\bc,k}^{\mathrm{H}}\bS_{\bc,k} + \langle \bLambda^{-1} \rangle \right)^{-1}. \label{posterior_var_alpha_k}
\end{align}

\textbf{2. Derivation of $q(\boldsymbol{\lambda})$}: Plugging in (\ref{prob_alpha_k}) and (\ref{prob_lambda}) to (\ref{VI_solution}), $\log q(\lambda_i)$ is derived by
\begin{align}
	\log q(\lambda_i) \propto -\frac{1}{2} \log \lambda_i - \frac{1}{2}\left( \frac{\langle \gamma_i \rangle}{2}\lambda_i + \frac{2\langle \vert \alpha_{k,i} \vert^2 \rangle}{\lambda_i} \right),
\end{align}
which suggests that $q(\lambda_i)$ follows a generalized inverse Gaussian distribution with the parameters $\frac{\langle \gamma_i \rangle}{2}$, $2 \langle \vert \alpha_{k,i} \vert^2 \rangle$, and $\frac{1}{2}$, and the posterior mean of $\lambda_i$ and its inverse are respectively given by \cite{Babacan:2014}
\begin{align} \label{posterior_mean_lambda} 
	\langle \lambda_i \rangle = \frac{2\sqrt{\langle \vert \alpha_{k,i} \vert^2 \rangle}}{\sqrt{\langle \gamma_i \rangle}} + \frac{2}{\langle \gamma_i \rangle}, \enspace
	\left \langle \lambda_i^{-1} \right \rangle = \frac{\sqrt{\langle \gamma_i \rangle}}{2 \sqrt{\langle \vert \alpha_{k,i} \vert^2 \rangle}}.
\end{align}

\textbf{3. Derivation of $q(\bgamma)$}: Plugging in (\ref{prob_lambda}) and (\ref{prob_gamma}) to (\ref{VI_solution}), $\log q(\gamma_i)$ is obtained by
\begin{align}
	\log q(\gamma_i) &\propto  \left(a+\frac{3}{2}-1\right)\log \gamma_i - \left(b +\frac{\langle \lambda_i \rangle}{4}\right) \gamma_i,
\end{align}
implying that $q(\gamma_i)$ follows a gamma distribution denoted by $\mathrm{Gamma}(\gamma_i \vert \bar{a}, \bar{b}_i)$ with
\begin{align}
	\bar{a} = a+\frac{3}{2}, \enspace \bar{b}_i = b+ \frac{\langle \lambda_i \rangle}{4},
\end{align}
and the posterior mean of $\gamma_i$ is
\begin{align} \label{posterior_mean_gamma}
	\langle \gamma_i \rangle = \frac{\bar{a}}{\bar{b}_i}.
\end{align}

\textbf{4. Derivation of $q(\beta)$}: Finally, plugging in (\ref{prob_y_k}) and (\ref{prob_beta}) to (\ref{VI_solution}), $\log q(\beta)$ is derived by 
\begin{align}
	\log q(\beta) &\propto \langle \log p(\by_k \vert \boldsymbol{\alpha}_{k},\beta)+\log p(\beta) \rangle_{-\beta} \nonumber
	\\ & \propto  (a+NT-1)\log\beta \nonumber
	\\ & \enspace \enspace \enspace -(b+\Vert \by_k-\bS_{\bc,k} \langle \boldsymbol{\alpha}_{k} \rangle\Vert_2^2 + \trace(\bC_{\boldsymbol{\alpha}_{k}}\bS_{\bc,k}^{\mathrm{H}}\bS_{\bc,k})
	)\beta.
\end{align}
This suggests that $q(\beta)$ follows a gamma distribution denoted by $\mathrm{Gamma}(\beta \vert \bar{a}_{\beta}, \bar{b}_{\beta})$, where its shape and rate parameters are respectively given by
\begin{align}
	\bar{a}_{\beta} &= a +NT, \nonumber \\
	\bar{b}_{\beta} &= b+\Vert \by_k-\bS_{\bc,k} \bm_{\boldsymbol{\alpha}_{k}} \Vert_2^2 + \trace(\bC_{\boldsymbol{\alpha}_{k}}\bS_{\bc,k}^{\mathrm{H}}\bS_{\bc,k}).
\end{align}
Thus, the posterior mean of $\beta$ is
\begin{align} \label{posterior_mean_beta}
	\langle \beta \rangle = \frac{\bar{a}_{\beta}}{\bar{b}_{\beta}}.
\end{align}

The proposed estimator discussed thus far is summarized in Algorithm 1, which sequentially updates the posterior means of the variables in $\Omega$ until convergence, and this sequential update is guaranteed to converge to a stationary point \cite{Bishop:2006}. The final output of the algorithm is $\bm_{\boldsymbol{\alpha}_{k}}$, from which the cascaded channel estimate is reconstructed by $\hat{\bc}_k = \bW_k \bm_{\boldsymbol{\alpha}_{k}}$. 

\begin{algorithm}[t]
	\textbf{Input:} $\by_k$, $\bS_{\bc,k}$ \enspace \textbf{Output:} $\bm_{\boldsymbol{\alpha}_{k}}$
	\begin{algorithmic} [1]
		\caption{Proposed Bayesian estimator}
		\State Set the parameters $a$, $b$ for the gamma hyperpriors
		\State Initialize $\bC_{\boldsymbol{\alpha}_{k}}$, $\langle \boldsymbol{\lambda} \rangle$, $\langle \boldsymbol{\lambda}^{-1} \rangle$, $\langle \bgamma \rangle$, and $\langle \beta \rangle$
		\While{termination condition}
			\State Update $\bm_{\boldsymbol{\alpha}_{k}}$ and $\bC_{\boldsymbol{\alpha}_{k}}$ according to (\ref{posterior_mean_alpha_k}) and (\ref{posterior_var_alpha_k})
			\State Update $\langle \boldsymbol{\lambda} \rangle$ and $\langle \boldsymbol{\lambda}^{-1} \rangle$ according to (\ref{posterior_mean_lambda})
			\State Update $\langle \bgamma \rangle$ according to (\ref{posterior_mean_gamma})
			\State Update $\langle \beta \rangle$ according to (\ref{posterior_mean_beta})
		\EndWhile
	\end{algorithmic}
\end{algorithm}

\section{Numerical Results} \label{sec_numerical}
In this section, we evaluate the performance of the proposed VI-based estimator. In the considered setup, the system consists of $N=16$ BS antennas, $K=3$ UEs, and a total of $L=10 \times 10$ RIS elements.
The BS and RIS are located at (0 m, 0 m) and (350 m, 10 m), while the UEs are uniformly distributed around a circle centered at (400 m, 0 m) with radius 5 m.
The uplink transmit power from the UEs is $P=\mbox{23 dBm}$. The noise variance at the BS is set as $\sigma_{\mathrm{B}}^2 = W \times N_0 \times \mathrm{NF}$, where $W=\mbox{80 MHz}$ is bandwidth, $N_0=-\mbox{174 dBm/Hz}$ is a noise spectral density, and $\mathrm{NF}=\mbox{7 dB}$ is noise figure.
The path-loss is modeled as $\mathrm{PL}= \mu_0(d/d_0)^{-\eta}$, where $\mu_0=-20$ dB is the reference path-loss at the reference distance $d_0 = 1$ m, $d$ is link distance in meters, and $\eta$ is the path-loss exponent. Specifically, the path-loss exponents for the RIS-BS and UE-RIS channels are set as $\eta_{\mathrm{RB}}=2.2$ and $\eta_{\mathrm{UR}}=2.1$, respectively.
The number of propagation paths is set to $M_{\mathrm{RB}}=2$ and $M_{\mathrm{UR},1}=\cdots=M_{\mathrm{UR},K}=3$. The number of time slots in each subblock is $\tau=K$. The shape and rate parameters for the gamma priors are set to $a=b=10^{-6}$.
The RIS phase shifts in $\bS$ are randomly sampled from a uniform distribution over the range of $[0, 2\pi)$.

As the performance metric, we will consider the following normalized mean squared error (NMSE) to quantify the cascaded channel estimation error:
\begin{align}
	\mathrm{NMSE}=\frac{1}{K} \sum_{k=1}^K \frac{\Vert  \bc_k - \hat{\bc}_k \Vert ^2}{\Vert \bc_k \Vert^2}.
\end{align}

\begin{figure}
	\centering
	\includegraphics[width=0.6\columnwidth]{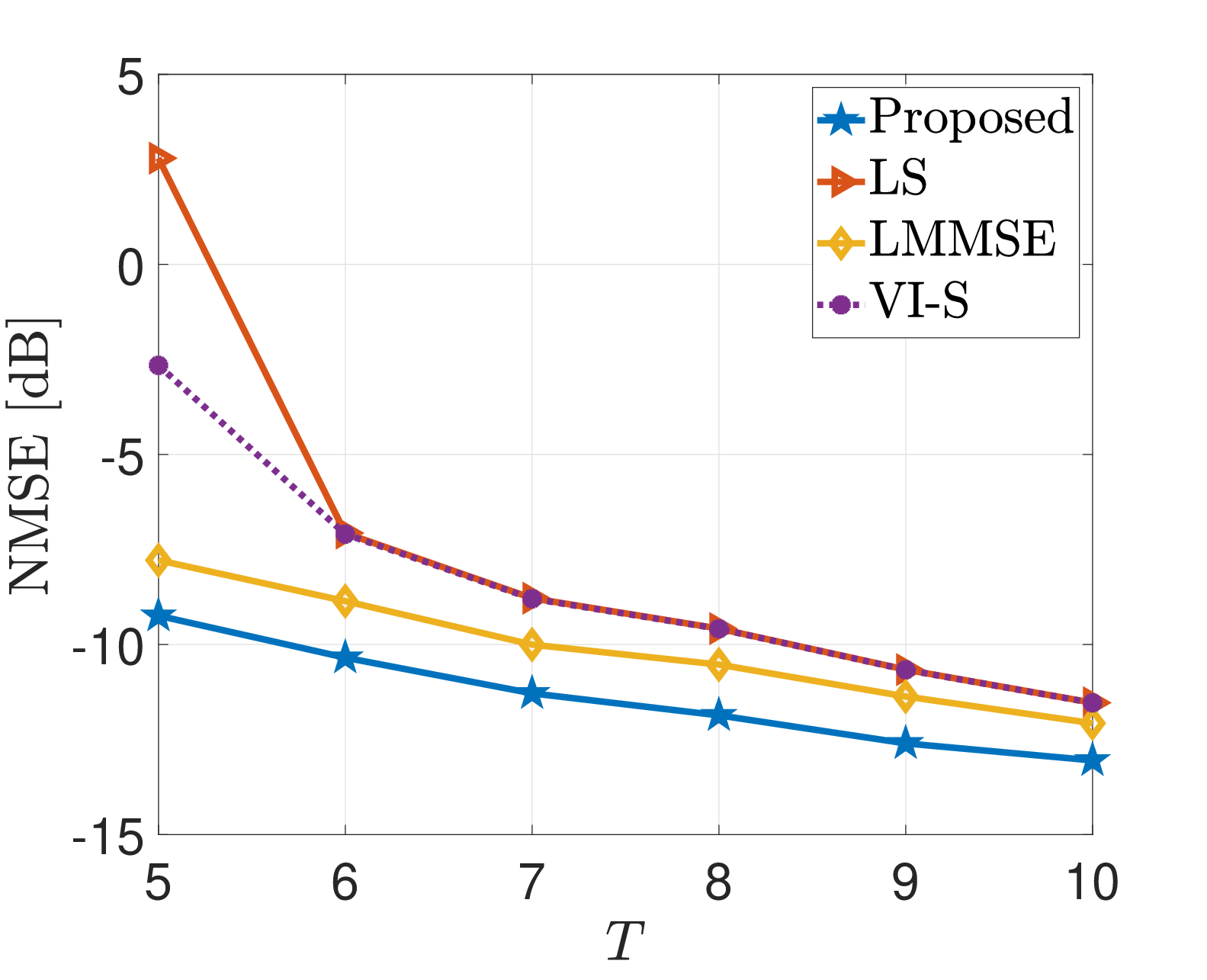}
	\caption{NMSE comparison versus number of blocks.}	\label{NMSE_vs_T}
\end{figure}

In Fig. \ref{NMSE_vs_T}, we compare the cascaded channel estimation performance according to the number of blocks $T$. As baselines, LS, LMMSE, and VI with a Student's-$t$ prior (VI-S) \cite{Tipping:2001} are considered.
Note that the computational complexities of all methods are $\cO(I N^2 L T M_k)$ for the proposed technique and the VI-S with the number of iterations $I$, $\cO(N^2 L T M_k)$ for the LS, and $\cO(N^3 L T^2)$ for the LMMSE.
It is observed that, the proposed technique achieves the lowest NMSE with the manageable complexity, indicating that an appropriate estimation technique for the complex path gains in the cascaded channel is necessary.
The performance gap between the proposed technique and both LMMSE and VI-S highlights that the complex adaptive Laplace distribution more accurately captures the PDF of the complex path gains in the cascaded channel compared to the Gaussian and Student's-$t$ distributions.

\begin{figure}
	\centering
	\includegraphics[width=0.6\columnwidth]{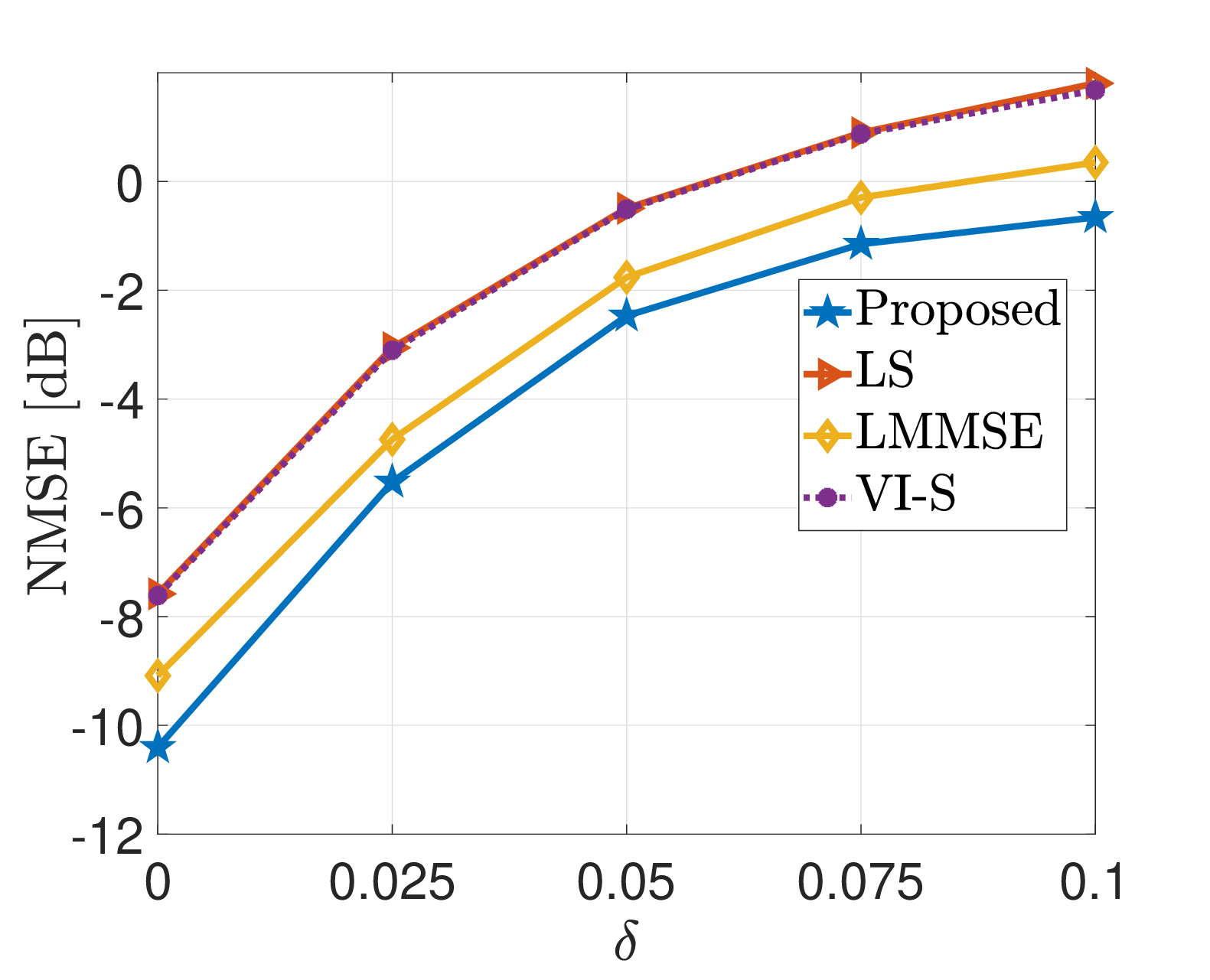}
	\caption{NMSE comparison under angle uncertainty.}	\label{NMSE_vs_delta}
\end{figure}

From Fig. \ref{NMSE_vs_delta}, we plot an NMSE comparison of the cascaded channel estimates with $T=6$ under angle uncertainty, where the estimation errors of the AoAs and AoDs in the RIS-related channels are modeled as zero-mean Gaussian distributions with variance $\delta^2$.
Similar to the results shown in Fig. \ref{NMSE_vs_T}, the proposed technique outperforms all considered baselines, with relatively stable performance gap as the uncertainty increases.
This demonstrates that, even under imperfect angle information cases, the proposed technique accurately estimates the complex path gains in the cascaded channel\footnote{We also verified that the proposed technique performs well in practical scenarios where angles of the RIS-related channels are estimated using a method such as the one proposed in \cite{Zhang:2021}.}.

\section{Conclusion} \label{conclusion}
In this paper, we have developed a Bayesian framework for estimating the cascaded channel in RIS-aided mmWave systems. To effectively approximate the complex channel gains in the cascaded channel, we employed a complex adaptive Laplace prior and applied VI to derive an approximate posterior distribution.
Numerical results demonstrated that the proposed estimator outperforms conventional estimators commonly used in systems without RISs.

\bibliographystyle{IEEEtran}
\bibliography{refs_all}

\end{document}